\documentclass[pra,preprint,showpacs]{revtex4}
\usepackage[centertags]{amsmath}
\usepackage{amsfonts}
\usepackage{amssymb}
\usepackage{amsthm} 
\usepackage{newlfont}
\usepackage{epsfig}
\usepackage{amscd}
\usepackage{graphicx}
\usepackage{epsfig}
\usepackage{footnote}
\usepackage{lipsum}
\usepackage{color}
\usepackage{xcolor}
\usepackage{amscd}

\newcommand{\beq}{\begin{equation}}
\newcommand{\eeq}{\end{equation}}
\newcommand{\ba}{\begin{array}}
\newcommand{\ea}{\end{array}}
\newcommand{\bea}{\begin{eqnarray}}
\newcommand{\eea}{\end{eqnarray}}
\begin{document}

\begin{center}
{\large \sc \bf {Remote two-qubit state creation and its robustness 
}}

\vskip 15pt

{\large 
J.Stolze and A.I.~Zenchuk 
}

\vskip 8pt

{\it Institute of Problems of Chemical Physics, RAS,
Chernogolovka, Moscow reg., 142432, Russia},\\
{\it Technische Universit\"at Dortmund, Fakult\"at Physik, D-44221 Dortmund, Germany}

\end{center}


\begin{abstract}
We consider the problem of remote two-qubit state creation using the two-qubit excitation pure
 initial state of the sender. The communication line is based  on the 
optimized boundary controlled chain with two pairs of properly adjusted coupling constants.
The creation of the two-qubit 
 Werner state  is considered as an  example. We also study the effects of imperfections of the chain on the state-creation.
 \end{abstract}

\maketitle

\section{Introduction}
\label{Sec:Introduction}

Information processing devices, be they classical or quantum in
nature, consist of storage elements, gates, and interconnections along
which information can be transferred. In a quantum device information
is contained in the quantum state of one or more quantum bits, or
qubits. For the transfer of information between different locations
several concepts have been developed. Quantum  state transfer 
\cite{Bose,CDEL,ACDE,KS,GKMT,WLKGGB,NJ,SAOZ} denotes the transport of
an existing state to a different place by the dynamics of an
underlying Hamiltonian with or without time-dependent control
parameters. In contrast, remote state creation
\cite{ZZHE,BPMEWZ,BBMHP,PBGWK2,PBGWK,DLMRKBPVZBW,XLYG} means the
preparation of a desired state in a desired location by manipulations
performed at an earlier time in a different location. Although this
technique was first realized in quantum optics 
\cite{PBGWK2,PBGWK,XLYG}, systems of
coupled spins have also been studied in this context 
 and may be used
for short-distance communication, for instance, within a quantum
information processing device. 

The principal setup for remote state creation used in our previous
papers \cite{Z_2014,BZ_2015,BZ_arxiv2015,DZ_arxiv2015,FKZ_arxiv2015}
is as follows. There are two small sets of qubits called sender and
receiver, respectively, connected by a chain of other qubits serving
as a transmission line. The state of the sender is characterized by a
set of parameters which may be adjusted in order to create the desired
state in the receiver subsystem. These are the control parameters. The
state of the receiver is also characterized by a set of parameters,
the target parameters. The possible states of the receiver occupy a
certain region in the space of target parameters. In recent papers
we have
studied various aspects of the remote creation of single-qubit states:
the creatable region \cite{Z_2014}, i.e. the region in the target parameter space
corresponding to states which can actually be remotely created by
adjusting the control parameters, the creation of states from pure
one-excitation initial states \cite{BZ_2015}, the enhancement of performance by
unitary transformations on the receiver side \cite{BZ_arxiv2015}, the remote creation of
quantum correlations (discord) \cite{DZ_arxiv2015} as well as the control of polarization
and coherence intensity of the created state \cite{FKZ_arxiv2015}. 

For transmission lines (spin chains) designed for perfect state
transfer \cite{CDEL,KS} the creatable region covers all possible
states of the receiver. However, these chains require careful
adjustment of all couplings along the chain which is difficult and
sensitive towards ``manufacturing errors'', small imperfections in the
coupling constants which reduce the perfect state transfer to a
high-probability state transfer. Since a high-probability state
transfer may also be achieved much more simply by boundary-controlled
\cite{GKMT,WLKGGB}  or  optimized boundary controlled chains
\cite{BACVV,ZO,BACV,ABCVV}, we study this type of systems in the present paper.

Our goal here is the remote creation of two-qubit states. For that
purpose we have to employ initial states with two excitations instead
of the one-excitation states studied previously. While the creatable
region for a one-qubit receiver is essentially (i.e. up to an
unimportant phase) defined by the values of two parameters and thus
can be shown in a single picture \cite{Z_2014,BZ_2015}, this becomes
much more complicated for a two-qubit receiver. There, the density
matrix depends on 15 parameters and thus is difficult to
visualize. Therefore we have to employ other means of controlling the
creatable {{}states. We show that the communication line 
is characterized by a finite number of Hamiltonian-dependent parameters 
and this number is defined by the dimensions of 
the sender and receiver  and does not depend on the whole length of the chain.}

The structure of this paper is as follows. In
Sec.\ref{Section:general} we present the general protocol of two-qubit state creation 
without specifying the interaction Hamiltonian. Then, in Sec.\ref{Section:model},
 we apply this protocol to 
 a particular  spin-1/2 
model governed by the XY Hamiltonian and describe the creation of the Werner state as an example. The effect of imperfections of the Hamiltonian on the state creation is considered
 in Sec. \ref{Section:imperfections}. Sec.\ref{Section:conclusions}
 contains concluding remarks .

\section{General protocol of remote two-qubit state creation}
\label{Section:general}
Our protocol
employs a $N$-node spin-1/2 chain 
and  
assumes that the evolution of this chain 
is governed by a Hamiltonian commuting with the $z$-projection of the total spin momentum $I_z$: $[H,I_z]=0$. 
Then the spin dynamics can be described in the subspace  spanned by the vectors 
\begin{eqnarray}\label{basis}
|0\rangle,\;\;|k\rangle, \;k=1,\dots,N,\;\; |nm\rangle, \;\;n=1,\dots,N-1,\;m=2,\dots,N,\;m>n,
\end{eqnarray}
whose dimensionality is $N+1 + \left({N}\atop{2}\right) = \frac{1}{2}(N^2+N+2)$.
In eqs.(\ref{basis}),  $|k\rangle$ means the state with the $k$th spin excited, $|nm\rangle$ means 
the state with the $n$th and $m$th spins excited, and $|0\rangle$ is the state without excitations.

\subsection{General form of the two-excitation initial state}
In this section we do not restrict the size  $N_S$ of the sender  and consider 
the general form of the two-qubit  initial state:
\begin{eqnarray}\label{inst}
|\Psi_0\rangle  = a_0 |0\rangle +  \sum_{i=1}^{N_S} a_i |i\rangle+  \sum_{{i,j=1}\atop{j> i}}^{N_S} 
a_{ij} |ij\rangle  ,
\end{eqnarray}
with normalization
\begin{eqnarray}\label{norm}
|a_0|^2 + \sum_{i=1}^{N_S} |a_i|^2 +  \sum_{{i,j=1}\atop{j> i}}^{N_S} 
|a_{ij}|^2 =1.
\end{eqnarray}
Here $a_i$ and  $a_{jk}$ are the amplitudes of the inital state.
The initial state $|\Psi_0\rangle$ evolves according to the 
Schr\"odinger equation as $|\Psi(t)\rangle = e^{- i H t}|\Psi_0\rangle$, where $H$ 
is the Hamiltonian governing the spin-dynamics.

\subsection{State of he two-qubit receiver and its characteristics}
\label{Section:receiver}
 The density matrix $\rho^R$ of the state of the two-qubit receiver is reducible from the density matrix 
  $|\Psi(t)\rangle \langle\Psi(t)|$ of the state of the whole system by tracing over all nodes except 
for the two nodes of the receiver. 
In the basis $|0\rangle$,  
$|N-1\rangle$, $|N\rangle$, $|(N-1)N\rangle$, the matrix $\rho^R$  reads:
\begin{eqnarray}\label{rhoR}
&&\rho^R(t) ={\mbox{Tr}}_{1,\dots,N-2} |\Psi(t)\rangle \langle\Psi(t)| =\\\nonumber
 &&
\left(
\begin{array}{cccc}
1-\rho_{N-1;N-1} -\rho_{N;N} -\rho_{(N-1)N;(N-1)N} &  \rho_{0;N-1}&  \rho_{0;N}&\rho_{0;(N-1)N}\cr
 \rho_{0;N-1}^*&\rho_{N-1;N-1} &  \rho_{N-1;N}&  \rho_{N-1;(N-1)N}\cr
 \rho_{0;N}^*& \rho_{N-1;N}^*& \rho_{N;N} &   \rho_{N;(N-1)N}\cr
\rho_{0;(N-1)N}^*& \rho_{N-1;(N-1)N}^*& \rho_{N;(N-1)N}^* &   \rho_{(N-1)N;(N-1)N}
\end{array}
\right).
\end{eqnarray}
In (\ref{rhoR}), all  elements $\rho_{ij}\equiv \rho_{ij}(t)$ 
are expressed in terms of the transition amplitudes  $f_i$ and $f_{ij}$
 from the inital state to {{}the} basis states:
\begin{eqnarray}\label{fi}
&&
f_i(t)=\langle i |\Psi(t)\rangle  = \sum_{k=1}^{N_S} a_k p_{i;k}(t),  
\\\label{fij}
&&
f_{ij}(t)=\langle ij |\Psi(t)\rangle =\sum_{{n,m=1}\atop{m> n}}^{N_S}  a_{nm} p_{ij;nm}(t)  ,
 \end{eqnarray}
where 
\begin{eqnarray}\label{p}
p_{i;k} = \langle i| e^{-i H t}|k\rangle, \quad
p_{ij;nm} = \langle ij| e^{-i H t}|nm\rangle 
\end{eqnarray}
are the transition amplitudes between the basis vectors.
In addition,  we  assume that the state $|0\rangle$ corresponds to  zero energy, 
hence  $f_0=\langle 0 |\Psi(t)\rangle=\langle 0 |\Psi_0\rangle = a_0$, {{}and $a_0$ is a real number}.
Notations (\ref{fi}) and (\ref{fij}) show that (i) the transition amplitudes $f_i$ and $f_{ij}$ are  linear functions of the initial state 
amplitudes $a_i$ and $a_{ij}$, and (ii) the dependence on the
Hamiltonian (in particular,  on the coupling constants  
characterizing the transmission line) is confined to the  parameters $p_{i;k}$ and $p_{ij;nm}$. Hence, these parameters
 can be considered fixed for a given transmission line while 
the amplitudes $a_i$ and $a_{ij}$ serve as control parameters of the sender which 
can be varied in order to create the desired state of the receiver. We collect these parameters in the list $a$:
\begin{eqnarray}\label{contrpar}
a=\{{{}a_0},a_k, a_{nm}: 
k= 1,\dots, N_S,\;n=1,\dots, N_S-1,\;m = 2,\dots, N_S,\;\;m>n \}.
\end{eqnarray}
{{}This list consists of $N_S^2+N_S$ independent  real parameters.}

The direct calculation of the elements of the density matrix $\rho^R$ yields: 
\begin{eqnarray}\label{2up1}
&& \rho_{0;N-1}=f_0 f_{N-1}^* +
\sum_{i=1}^{N-2}f_{i} f_{i(N-1)}^*=a_0 \sum_{k=1}^{N_S}p_{N-1;k}^* a_k^*+
\sum_{{k,n,m=1}\atop{m>n}}^{N_S} P_{N-1;knm} a_k a_{nm}^*, 
\\
&&
\rho_{0;N}=f_0 f_{N}^* +
\sum_{i=1}^{N-2}f_{i} f_{iN}^*=a_0 \sum_{k=1}^{N_S}p_{N;k}^* a_k^*+
\sum_{{k,n,m=1}\atop{m>n}}^{N_S} P_{N;knm} a_k a_{nm}^* ,
\\\label{2up12}
&& \rho_{0;(N-1)N}=f_0 f_{(N-1)N}^*=
a_0\sum_{{n,m=1}\atop{m>n}}^{N_S} p_{(N-1)N;nm}^*  a_{nm}^*,
\end{eqnarray}
\begin{eqnarray}
\label{2up2}
&& \rho_{N-1;N-1}=f_{N-1} f_{N-1}^*+ \sum_{i=1}^{N-2}f_{i(N-1)} f_{i(N-1)}^* =\\\nonumber
&&
\sum_{n,m=1}^{N_S} p_{N-1;n} p_{N-1;m}^* a_n a_m^* +
\sum_{{k,l,n,m=1}\atop{m>n,l>k}}^{N_S} P_{(N-1)(N-1);klnm} a_{kl} a_{nm}^* 
,\\\label{2up23} 
&&
\rho_{N-1;N}=f_{N-1} f_{N}^* + \sum_{i=1}^{N-2}f_{i(N-1)} f_{iN}^* =\\\nonumber
&&
\sum_{n,m=1}^{N_S} p_{N-1;n} p_{N;m}^* a_n a_m^* +
\sum_{{k,l,n,m=1}\atop{m>n,l>k}}^{N_S} P_{(N-1)N;klnm} a_{kl} a_{nm}^* 
, \\\label{2up22} 
&&\rho_{N-1;(N-1)N}=f_{N-1} f_{(N-1)N}^*=
\sum_{{k,n,m=1}\atop{m>n}}^{N_S} p_{N-1;k} p_{(N-1)N;nm}^* a_k a_{nm}^* ,
\end{eqnarray}
\begin{eqnarray}
\label{2up3}
&& \rho_{N;N}=f_{N} f_{N}^* + \sum_{i=1}^{N-2}f_{iN} f_{iN}^*
=
\sum_{n,m=1}^{N_S} p_{N;n} p_{N;m}^* a_n a_m^* +
\sum_{{k,l,n,m=1}\atop{m>n,l>k}}^{N_S} P_{NN;klnm} a_{kl} a_{nm}^* 
, \\\label{2up32} 
&&\rho_{N;(N-1)N}=f_{N} f_{(N-1)N}^*=
\sum_{{k,n,m=1}\atop{m>n}}^{N_S} p_{N;k} p_{(N-1)N;nm}^* a_k a_{nm}^* ,\\\label{2up4}
&&
\rho_{(N-1)N;(N-1)N}=f_{(N-1)N} f_{(N-1)N}^* =
\sum_{{k,l,n,m=1}\atop{l>k,m>n}}^{N_S} p_{(N-1)N;kl} p_{(N-1)N;nm}^* a_{kl} a_{nm}^* . 
\end{eqnarray}
Here we introduce the following notations:
\begin{eqnarray}\label{Pparam}
&&
P_{N-1;knm} =\sum_{i=1}^{N-2} p_{i;k} p_{i(N-1);nm}^*  ,
\;\;
P_{N;knm} =\sum_{i=1}^{N-2} p_{i;k} p_{iN;nm}^*,\\\nonumber
&&
P_{(N-1)(N-1);klnm} =\sum_{i=1}^{N-2} p_{i(N-1);kl} p_{i(N-1);nm}^*
,\\\nonumber
&&
P_{(N-1)N;klnm} =\sum_{i=1}^{N-2} p_{i(N-1);kl} p_{iN;nm}^*
,
\;\;P_{NN;klnm} = \sum_{i=1}^{N-2} p_{iN;kl} p_{iN;nm}^*.
\end{eqnarray}
{{}Formulas (\ref{2up1}-\ref{2up4}) show that the receiver's density matrix 
depends on the set of parameters}
\begin{eqnarray} \label{char}\label{param}
{\cal{P}}&=&\{
p_{N;i}, \;\; p_{N-1;i}, \;\;p_{(N-1)N;nm},\;\; P_{N-1;inm},\;\; P_{N;inm},\;\; \\\nonumber
&&
P_{(N-1)(N-1);klnm}, \;\;
P_{(N-1)N;klnm}, \;\;
P_{NN;klnm}, \\\nonumber
&& i=1,\dots, N_S,\;\; k,n=1,\dots,N_S-1,\;\; l,m=2,\dots,N_S, \;\;k>l,\;\; m>n\},
\end{eqnarray}
which represent {{} $\frac{1}{4} (3 N_S^2 - 5 N_S +6)(N_S+1)N_S$  complex characteristics of the transmission
line  depending} on the particular Hamiltonian (governing the spin
dynamics), the  length of the communication 
line   and  the time instant $t$.
There is the following symmetry among these parameters:
\begin{eqnarray} \label{symm}
P_{NN;klnm} = P_{NN;nmkl}^*,\;\;P_{(N-1)(N-1);klnm} = P_{(N-1)(N-1);nmkl}^*.
\end{eqnarray}
All parameters in ${\cal{P}}$ can be directly defined for a given chain at any preferred  time instant $t=t_0$
(see Sec.\ref{Section:dp}) and do not change during  operation of the communication line.

It is important that the number of these parameters depends only
on the dimensionality of the sender and receiver and does not depend on the length of 
the communication line. Below, for our convenience we use the notation $P$ without subscript 
 for a general element of the set 
${\cal{P}}$: $P\in {\cal{P}}$.

\subsection{Defining  the characteristics (\ref{param}) of  the communication line}
\label{Section:dp}
Although there are exact formulas (\ref{p}) and (\ref{Pparam}) involving the Hamiltonian $H$ and the time $t$,
 the actual values of these parameters for a given communication line 
can differ from the analytically calculated ones because of
imperfections in the Hamiltonian. 
Therefore, before proceeding to
operating a particular communication line, 
we have to define its  parameters ${\cal{P}}$.  For this purpose we create a set of preliminary states 
of the receiver using the following set of specially selected pure
inital states of the sender:
%
\begin{eqnarray}\label{1n}
&&
a_0 |0\rangle + a_i |i\rangle,\;\;a_0^2+a_k^2=1,
\\\label{3n}
&&
a_i |i\rangle + a_{nm} |nm\rangle,\;\;a_k^2+a_{nm}^2=1, \;\;m> n\\\label{4n}
&&
a_{kl} |kl\rangle + a_{nm} |nm\rangle,\;\;a_{kl}^2+a_{nm}^2=1,\;\;m> n,\;\; l> k,\\\label{5n}
&&
a_{kl} |kl\rangle + i a_{nm} |nm\rangle,\;\;a_{kl}^2+a_{nm}^2=1,\;\;m> n,\;\; l> k,
\\\nonumber
&& i=1,\dots,N_S,\;\;\;\; 
 \;\;l,m=2,\dots,N_S,\;\;k,n=1,\dots,N_S-1,
\end{eqnarray}
with some known values of the real constants $a_k$ and $a_{nm}$. Then,
the resulting  states of the  receiver will provide us with
information about  the  set of  parameters  ${\cal{P}}$. 
This process can be considered the solution of  a direct problem. On the contrary,
finding the set of 
control parameters needed 
for the creation of a desirable receiver's state is an inverse problem and therefore is more tricky. 

Now we proceed to defining the set of parameters ${\cal{P}}$ using the formulas (\ref{2up1}-\ref{2up4}) 
for the elements of the density matrix (\ref{rhoR}) and the   above set of initial conditions 
 (\ref{1n}-\ref{5n}) at the prescribed time instant $t=t_0$ (the way of  fixing this time instant 
will be discussed in Sec.\ref{Section:model}). 

\noindent
Initial condition (\ref{1n}) yields {{}(we give only the nonzero entries of the density matrix $\rho^R$ in formulas 
(\ref{2up1in1}-\ref{2up1in4}))}
\begin{eqnarray}\label{2up1in1}
&& \rho_{0;N-1}=a_0 p_{N-1;k}^* a_k, 
  \;\;
\rho_{0,N}=a_0 p_{N,k}^* a_k ,
,\;\; \rho_{N-1;N-1}= |p_{N-1;k}|^2 a_k^2 
,\\\nonumber
&&
\rho_{N-1;N}= p_{N-1;k} p_{N;k}^* a_k^2  
, \;\;
\rho_{N;N}=|p_{N;k}|^2 a_k^2  ,
\end{eqnarray}
defining $p_{N;k}$ and $p_{N-1;k}$. 

\noindent
Initial condition (\ref{3n}) yields
\begin{eqnarray}\label{2up1in3}
&& \rho_{0;N-1}=
 P_{N-1;knm} a_k a_{nm}, 
 \;\;
\rho_{0;N}= P_{N;knm} a_k a_{nm} ,
\\\nonumber
&& \rho_{N-1;N-1}= |p_{N-1;k}|^2 a_k^2 +P_{(N-1)(N-1);nmnm}  a_{nm}^2
,\;\;
\rho_{N-1;N}= p_{N-1;k} p_{N;k}^* a_k^2 +
 P_{(N-1)N;nmnm}  a_{nm}^2
, \\\nonumber
&&\rho_{N-1;(N-1)N}= p_{N-1;k} p_{(N-1)N;nm}^* a_k a_{nm} ,
\;\;
\rho_{N;N}= |p_{N;k}|^2 |a_k|^2 + P_{NN;nmnm} a_{nm}^2 
, \\\nonumber 
&&\rho_{N;(N-1)N}= p_{N;k} p_{(N-1)N;nm} a_k a_{nm} ,
\;\;\rho_{(N-1)N;(N-1)N}=|p_{(N-1)N;nm}|^2  a_{nm}^2,
\end{eqnarray}
 defining  {{}$p_{(N-1)N;nm}$, $P_{(N-1)N;nmnm}$, $P_{(N-1)(N-1);nmnm}$, $P_{NN;nmnm}$, $P_{N-1;knm}$ and $P_{N;knm}$.}
 
\noindent
Finally, initial condition (\ref{4n}) yields, in virtue of symmetry (\ref{symm}),
\begin{eqnarray}\label{2up1in4}
 \rho_{N-1;N-1}&=&
 P_{(N-1)(N-1);nmnm} a_{nm}^2+  P_{(N-1)(N-1);klkl} a_{kl}^2 + \\\nonumber
&&
2{\mbox{Re}}(P_{(N-1)(N-1);klnm}) a_{kl}a_{nm}
,\\\label{2up23dp} \nonumber
\rho_{N-1;N}&=&
 P_{(N-1)N;nmnm} a_{nm}^2+  P_{(N-1)N;klkl} |a_{kl}|^2 + \\\nonumber
&&
2{\mbox{Re}}(P_{(N-1)N;klnm})  a_{kl} a_{nm}
, \\\label{2up22dp} \nonumber
 \rho_{N;N}&=&
 P_{NN;nmnm} a_{nm}^2+  P_{NN;klkl} a_{kl}^2 + 
2{\mbox{Re}}(P_{NN;klnm})  a_{kl} a_{nm}
, \\\label{2up32dp}\nonumber
\rho_{(N-1)N;(N-1)N}&=&  |p_{(N-1)N;nm}|^2 a_{nm}^2+|p_{(N-1)N;kl}|^2 a_{kl}^2
 +\\\nonumber
&&2{\mbox{Re}}(p_{(N-1)N;kl} p_{(N-1)N;nm}^*) a_{kl} a_{nm},
\end{eqnarray}
defining the real parts of  $P_{(N-1)(N-1);klnm}$, $P_{(N-1)N;klnm}$ and $P_{NN;klnm}$, $(kl)\neq (nm)$.
The imaginary parts of these parameters can be obtained using the initial condition  (\ref{5n}).

As a result, we obtain the whole list of parameters  ${\cal{P}}$ (\ref{param}) for a given communication line at 
the prescribed time instant $t_0$.

\subsection{Creation of  desired state. Approximate states. }
\label{Section:creation}
In order to create the desired state $A$ of the receiver, we have to solve the following system of algebraic quadratic 
 equations  (the inverse problem):
\begin{eqnarray}\label{rhoA}
\rho^R_{ij}({{}{\cal{P}}},a) = A_{ij},\;\;i,j=1,\dots,4, j\ge i,
\end{eqnarray}
for the control parameters $a_i$, $a_{ij}$
subject to the normalization condition (\ref{norm}). 
Here the elements of the {{}receiver's  density} matrix  $\rho^R$  are  given by formulas (\ref{2up1}-\ref{2up4}) 
and $a$ is a list of all control parameters (\ref{contrpar}).
If a solution of system (\ref{rhoA},\ref{norm}) exists, then the state $A$ may be exactly 
created using our communication line. 
However, in many cases we cannot create the state $A$ exactly even if the system 
 (\ref{rhoA},\ref{norm}) is solvable. This may happen due to any of
 the following reasons.
\begin{enumerate}
\item
{{}The  set of parameters ${\cal{P}}$ cannot be exactly defined {{} in general},  see Sec.\ref{Section:param}.
So,  system (\ref{rhoA}) must be replaced with the following one\footnote{Obviously, the matrix 
$\rho^R_{ij}({\cal{P}}^{apr},a)$ in eq.(\ref{rhoAapr})in
  general is not a density matrix. However, this is not important, because this system serves 
only to define the parameters $a$ which will be implemented in the
inital state of the sender. Then the matrix
 $\rho^R_{ij}({\cal{P}},a)$ obtained after evolution of the
 implemented initial state is a correct density matrix by
 construction, and
this matrix is used in formula (\ref{def}) introducing the matrix discrepancy.}:
\begin{eqnarray}\label{rhoAapr}
\rho^R_{ij}({{}{\cal{P}}}^{apr},a) = A_{ij},\;\;i,j=1,\dots,4, j\ge i,
\end{eqnarray}
with approximate parameters ${{}{\cal{P}}}^{apr}$.}
\item
{{}The set of control parameters   $a$ found as a solution to system  (\ref{rhoAapr},\ref{norm}) 
cannot be 
exactly implemented in the experiment, see Sec.\ref{Section:Werner}. We implement the set $a^{apr}$ which approximates the set of 
 ideal parameters $a$ up to some accuracy}
\item
The Hamiltonian contains imperfections
(for instance, the coupling constants between nodes differ
 from the anticipated values), see  Sec.\ref{Section:imperfections}. 
\end{enumerate}
Thus we have to estimate the difference between the matrix {{}$\rho^R({\cal{P}}, a^{apr})$}
{{}(in this formula,  ${\cal{P}}$ is the  set of exact parameters because $\rho^R$
 is the matrix created in the experiment)}   and the desired matrix  $A$. For this purpose we introduce the matrix discrepancy
 $\delta(\rho^R({\cal{P}}, a^{apr}))$ by the following formula:
\begin{eqnarray}\label{def}
\delta(\rho^R({\cal{P}},a^{apr})) = \frac{||\rho^R({\cal{P}},a^{apr}) - A|| }{||A||},
\end{eqnarray}
where we use the standard matrix norm defined for a matrix $x$ as follows
\begin{eqnarray}
||x|| = \sqrt{\sum_{i,j=1}^N |x_{ij}|^2}
\end{eqnarray}
If $\delta(\rho^R({\cal{P}},a^{apr})) < \varepsilon \ll 1$, then we say that the desired state $A$ is created with 
the accuracy  $\varepsilon$.

\section{A particular spin model}
\label{Section:model}
\subsection{Interaction Hamiltonian}
\label{Section:ham}
The parameters (\ref{param}) described  in the previous section depend on the Hamiltonian governing the 
dynamics of a spin chain. 
Although the best result (the largest creatable space)  is anticipated for the chain 
engineered for the perfect state transfer, we 
base our study on a simpler boundary-controlled model described by the nearest-neighbor XY Hamiltonian 
with two  end-pairs of properly adjusted coupling constants:
 \begin{eqnarray}\label{XY}
H=&&  \sum_{i=3}^{N-3}D (I_{ix} I_{(i+1)x} + I_{iy} I_{(i+1)y}) +\\\nonumber
&& 
\delta_1 (I_{1x} I_{2x} + I_{1y} I_{2y} + I_{(N-1)x} I_{Nx} + I_{(N-1)y} I_{Ny})+\\\nonumber
&& 
\delta_2 (I_{2x} I_{3x} + I_{2y} I_{3y} + I_{(N-2)x} I_{(N-1)x} + I_{(N-2)y} I_{(N-1)y}),
\end{eqnarray}
where  $D$, $\delta_1$ and $\delta_2$  are  the coupling constants between the 
nearest neighbors, $I_{j\alpha}$ ($j=1,\dots,N$, $\alpha=x,y,z$) is the 
$j$th spin projection on the $\alpha$-axis. Below we put $D=1$ which corresponds to using dimensionless time.
The values of the coupling constants $\delta_i$, $i=1,2$, are chosen in such a way that 
 they  maximize  the probability $|p_{N;1}(t_0)|$ at some time instant $t_0$, i.e. they maximize the 
value of the first maximum of $|p_{N;1}(t_0)|$.
For instance, in the case of  $N=20,\;60$  we have  the following values of  
the parameters $\delta_i$, of the maximized 
 amplitude  $|p_{N;1}(t_0)|$ and of the corresponding time instant $t_0$:  
\begin{eqnarray}\label{dd}
N=20:&& \;\;\delta_1= 0.550,\;\;\delta_2=0.817,\;\;|p_{20;1}|=0.99606,\;\;t_0=26.441,\\\nonumber
N=60:&& \;\;\delta_1 =0.414,\;\;\delta_2 =0.720, \;\;|p_{60;1}|=0.99223,\;\; t_0=70.203.
\end{eqnarray}
Formulas (\ref{dd}) complete the definition of the Hamiltonian (\ref{XY}) and allow us to uniquely 
define the set of parameters ${\cal{P}}$ (\ref{param}).
We chose the dimensionality of the sender by the requirement that the number of real control parameters 
is not less than the number of 
creatable real  parameters which is 15 for the two-qubit receiver. Thus, the minimal dimensionality of the sender is 
$N_S=4$ (20 independent real control parameters in initial state (\ref{inst})) which is used hereafter.

\subsection{Classification  of  parameters   (\ref{param}) by absolute value }
\label{Section:param}
Parameters (\ref{param}) differ by their absolute values {{}
 which are determined by  two main 
factors. First, the mirror symmetry in $p_{i;j}$, $p_{kl;nm}$ appearing in the definitions (\ref{Pparam})
of parameters ${\cal{P}}$. 
Due to the spatial symmetry of the chain the
  transition amplitudes (\ref{p}) tend to be large when the sites or
  pairs of sites involved are at positions symmetric with respect to
  the center of the chain.
Second, the position of nodes $i$ and  $j$ in $p_{i;j}$ (or
$k$, $l$, $n$ and $m$ in $p_{kl;nm}$): these quantities decrease with approaching the 
center of  communication line. Thus, we separate the parameters into three families.}

{\bf Family I} consists of 13 parameters. For  chains of $N=20$ and $60$  nodes the  
absolute values of these parameters 
are  in the intervals   {{}
$(0.9356,0.9961)$ and  $(0.8341, 0.9923$), respectively,}
see Table \ref{Table1}. 
The large values of these parameters are explained by the mirror symmetry of 
the transition amplitudes $p_{i;i}$ and $p_{nm;nm}$ which appear in  definitions 
(\ref{Pparam}) of the parameters 
${\cal{P}}$. {{}For instance, the parameters $P_{19;3,2,3}$ and  $P_{19,20;2,3,1,3}$ 
contain the large terms  $p_{18;3} p_{19,18;2,3}$
and $p_{19,18;2,3} p_{20,18;1,3}$, respectively, and so on}.

 \begin{table}
\begin{minipage}{8cm}
\vspace*{0.9cm }
\begin{tabular}{|c|c|c|c|}
\hline
$P\in {\cal{P}}$&${\cal{N}}$&$N= 20$&$N= 60$\cr 
\hline
$p_{N-1;2}$&1&$0.96743 i$& $0.91422 i$\cr
$p_{N;1}$&2&$ 0.99606 i $&$ 0.99223 i$\cr
$p_{(N-1)N;1,2}$&3&$0.96361 $&$  0.90711 $ \cr 
$P_{N-1;3,2,3}$&4 &$0.96707 i $&$ 0.91238 i$\cr
$P_{N-1;4,2,4}$ &5&$0.96741 i $&$ 0.91422 i$\cr
$ P_{N;3,1,3}$&6&$0.99601 i $&$ 0.99220 i$\cr
$P_{N;4,1,4}$&7&$0.98812 i$&$ 0.94575 i$\cr
$P_{NN;1,3,1,3}$&8&$ 0.99246$&$  0.98649$\cr
$P_{NN;1,4,1,4} $&9&$  0.98424 $&$ 0.93841  $\cr 
$P_{(N-1)(N-1);2,3,2,3}$&10&$ 0.93561$& $ 0.83415$\cr
$P_{(N-1)(N-1);2,4,2,4}$&11 &$ 0.94387 $&$ 0.88263 $\cr
$P_{(N-1)N;2,3,1,3}$ &12&$0.96361$&$ 0.90711  $\cr
$P_{(N-1)N;2,4,1,4} $&13&$0.96361$&$ 0.90711$
\cr
\hline
\end{tabular}\caption{{{}Family I of parameters ${\cal{P}}$ (\ref{param});
$0.9356 <|P|<0.9961$ and   $0.8341 <|P|< 0.9923$ for  chains of 
$N=20$ and
 $N=60$ nodes, respectively.} 
   \label{Table1}}
\end{minipage}
\hfill
\begin{minipage}{8cm}
\begin{tabular}{|c|c|c|c|}
\hline
$P\in {\cal{P}}$&${\cal{N}}$&$N= 20$&$N= 60$\cr 
\hline
$p_{N-1;4} $&1&$- 0.08929 i$&$-0.21641i$ \cr
$p_{(N-1)N;1, 4}$&2&$-0.08894$&$-0.21473  $\cr
$P_{N-1;2, 2, 4}$&3&$0.08929 i$&$0.21641 i $ \cr
$P_{N-1;3, 3, 4}$&4&$0.08925 i$&$0.21598 i$ \cr
$P_{N;2, 1, 2} $&5&$0.06384 i$&$0.16292 i $\cr
$P_{N;4, 1, 2}$&6&$0.08604 i$&$0.19631 i $\cr
$P_{N;2, 1, 4} $&7&$0.08604 i$&$0.19631 i $\cr
$P_{NN;1, 2, 1, 2}$&8&$0.06358$&$0.16166$ \cr
$P_{NN;1, 4, 1, 2} $&9&$0.08570$&$0.19479$ \cr
$P_{NN;1, 2, 1, 4}$&10&$0.08570$&$0.19479 $\cr
$P_{(N-1)(N-1);3, 4, 2, 3} $&11&$0.08635$&$0.19745 $\cr
$P_{(N-1)(N-1);2, 3, 3, 4}$&12&$0.08635$&$0.19745 $\cr
$P_{(N-1)N;2, 4, 1, 2}$&13&$0.08894$&$0.21473 $\cr
$ P_{(N-1)N;3, 4, 1, 3} $&14&$0.08894$&$0.21473 $\cr
\hline
\end{tabular}
\caption{{{}Family II of parameters ${\cal{P}}$ (\ref{param});
$0.0635 <|P|<0.0893$ and   $0.1616 <|P|<0.2165$ for  chains of 
$N=20$ and
 $N=60$ nodes, respectively.}\label{Table21}}
\end{minipage}
\end{table}

{\bf Family II} consists of   14 parameters.
For the chains of $N=20$ and 60 nodes the  absolute values of these parameters 
are  in the intervals  {{}
 $(0.0635,0.0893)$ and $(0.1616,0.2165)$, respectively,}  see Table
 \ref{Table21}. The mirror symmetry just mentioned
is significantly reduced in these parameters {{}(in definitions   (\ref{Pparam}) of these parameters,
 there is no term consisting of a product of two symmetrical transition amplitudes).

The values of parameters ${\cal{P}}$ are given with the accuracy $\sim 10^{-5}$ in both tables.}
{{}The symmetry (\ref{symm}) is confirmed in Tables \ref{Table1} and \ref{Table21}. Equalities between some other elements
in this table are approximate and disappear in higher order approximations.}

{\bf Family III} consists of 143 parameters with destroyed mirror  symmetry and with 
 absolute values below{{} $0.0193$
 (for the chain $N=20$) and below 
$0.0468$ ($N=60$). }
This family for the chain of 20 nodes is   given in Appendix, Sec.\ref{Section:Appendix}. 

By  
adjusting all of the coupling constants in the Hamiltonian (\ref{XY}), 
we may   approach the chain engineered for 
perfect state transfer. In this limiting case  the
  transition amplitudes (\ref{p}) are equal to unity if $k=N+1-i$
  ($(n,m)=(N+1-i,N+1-j)$) \cite{KS} and thus all parameters from
  Family I tend
 to unity, while all  
other parameters vanish.

\subsection{Example of remote state creation:  Werner  state}
\label{Section:Werner}

As a particular example {{}we use a chain of 20 nodes to create} the Werner state \cite{Werner} which  in the two-qubit case reads as follows:
\begin{eqnarray}
\rho^W=\left(\begin{array}{cccc}
\frac{1-p}{4}&0&0&0\cr
0&\frac{1+p}{4}&-\frac{p}{2}&0\cr
0&-\frac{p}{2}&\frac{1+p}{4}&0\cr
0&0&0&\frac{1-p}{4}
\end{array}\right), \;\;0\le p \le 1.
\end{eqnarray}
To create {{}the zero entries in this matrix  with our tool of four node sender  we
have to  put  $a_i=0$,  $i=0,1,\dots 4$.} 
The system of equations {{}(\ref{rhoAapr})} for the control parameters $a_{ij}$ can be written equating
 the corresponding 
elements of the density matrices {{}
$\rho^R({\cal{P}}^{apr},a)$ and $\rho^W$ (we use the parameters ${\cal{P}}^{apr}$ from the Tables \ref{Table1} and
\ref{Table21} and from the Appendix)}:
\begin{eqnarray}\label{eq1}
&&
\sum_{{k,l,n,m=1}\atop{l>k,m>n}}^{N_S} p_{(N-1)N;kl} p_{(N-1)N;nm}^* a_{kl} a_{nm}^*=
\frac{1-p}{4},\\\label{eq2}
&&
\sum_{{k,l,n,m=1}\atop{m>n,l>k}}^{N_S} P_{(N-1)(N-1);klnm} a_{kl} a_{nm}^* =
\sum_{{k,l,n,m=1}\atop{m>n,l>k}}^{N_S} P_{NN;klnm} a_{kl} a_{nm}^*=\frac{1+p}{4},\\\label{eq3}
&&
\sum_{{k,l,n,m=1}\atop{m>n,l>k}}^{N_S} P_{(N-1)N;klnm} a_{kl} a_{nm}^* =-\frac{p}{2}.
\end{eqnarray}
In turn, eq. (\ref{eq3}) splits into the {{}real and imaginary} parts:
\begin{eqnarray}\label{eq4}
&&
{\mbox{Im}}\left(\sum_{{k,l,n,m=1}\atop{m>n,l>k}}^{N_S} P_{(N-1)N;klnm} a_{kl} a_{nm}^*\right) = 0,
\;\; {\mbox{Re}}\left(\sum_{{k,l,n,m=1}\atop{m>n,l>k}}^{N_S} P_{(N-1)N;klnm} a_{kl} a_{nm}^*\right) =-\frac{p}{2}.
\end{eqnarray}
In addition, we have the normalization condition 
(\ref{norm}) for the control parameters which now reads:
\begin{eqnarray}\label{eq6}
 \sum_{{i,j=1}\atop{j> i}}^{4} 
|a_{ij}|^2 =1.
\end{eqnarray}
{{} All in all, we have a system of 
 6 real equations  
(\ref{eq1},\ref{eq2},\ref{eq4},\ref{eq6}), so that we need 6 real control parameters to solve it. 
Consequently, since we are interested in particular solutions, we may consider only 
the real parameters $a_{ij}$.  The numerical investigation of this system shows that 
it is solvable for  $a_{ij}$ if 
 $p$ is  in the interval 
$0\le p \le 0.8744$.}
We collect the results of the calculations for the
  parameter values $p=0.1 n,$ $n=0,1\dots,8$, in Table \ref{Table0} to five digit accuracy
(the same accuracy is used for the parameters ${\cal{P}}$ in 
Sec.\ref{Section:param}). The appropriate discrepancies $\delta(\rho^R)$ are given in the last column.
\begin{table}
\begin{tabular}{|c|c|c|c|c|c|c|c|}
\hline
$p$&$a_{12}$&$a_{13}$&  $a_{14}$&$a_{23}$&$a_{24}$&$a_{34}$&$\delta(\rho^R)$\cr
\hline
0  & 0.53233&0.42337&0.20146&0.21791&-0.44593&0.50046&$4.235 \times 10^{-5}$\cr
0.1  & 0.52208&0.24658&0.41283&0.38198&-0.34211&0.48296&$2.641 \times 10^{-5}$\cr
0.2  & 0.49965&0.14204&0.48397&0.42820&-0.32433&0.45540&$1.532\times 10^{-5}$\cr
0.3  & 0.47309&0.07861&0.52371&0.44462&-0.34473&0.42334&$3.883 \times 10^{-5}$\cr
0.4  & 0.44366&0.03642&0.55344&0.44645&-0.38257&0.38713&$1.744\times 10^{-6}$\cr
0.5  & 0.41131&0.01047&0.57913&0.43460&-0.43236&0.34571&$2.066\times 10^{-5}$\cr
0.6  & 0.37550&0.00361&0.60326&0.40331&-0.49433&0.29673&$2.347\times 10^{-5}$\cr
0.7  & 0.33507&0.02962&0.62617&0.33516&-0.57203&0.23496&$6.332\times 10^{-6}$\cr
0.8  & 0.28714&0.13374&0.63657&0.17462&-0.66556&0.14487&$1.604\times 10^{-5}$\cr
 \hline
\end{tabular}
\caption{ A particular solution of  system (\ref{eq1},\ref{eq2},\ref{eq4},\ref{eq6}) with the accuracy $10^{-5}$ for different $p$,  
all parameters ${\cal{P}}$ are  taken into account  \label{Table0}. The last column shows the values of the 
discrepancy $\delta(\rho^R(a))$ between the created state and the Werner state.}
\end{table}

There is one more simplification of the model which follows from the fact that 
the absolute values  of the  parameters ${\cal{P}}$ from Family III 
are small and thus we can put all of them equal to zero {{} as the first approximation}. Solving the system  (\ref{eq1},\ref{eq2},\ref{eq4},\ref{eq6}) 
 for  $a_{ij}$ in this case we obtain new values for the control parameters collected in  
 in Table \ref{Table01} together with appropriate discrepancies $\delta(\rho^R)$ 
(the last  column).
\begin{table}
\begin{tabular}{|c|c|c|c|c|c|c|c|}
\hline
$p$&$a_{12}$&$a_{13}$&  $a_{14}$&$a_{23}$&$a_{24}$&$a_{34}$&$\delta(\rho^R)$\cr
\hline
0   & 0.55924&0.00127&0.43728&0.47083&-0.00135&0.52378&$2.906 \times 10^{-2}$\cr
0.1  &0.53510&-0.04674&0.46417&0.49567&-0.05174&0.49766&$2.893 \times 10^{-2}$\cr
0.2  &0.50894&-0.09432&0.48584&0.51641&-0.09569&0.46926&$2.760 \times 10^{-2}$\cr
0.3  &0.48066&-0.13718&0.50409&0.53327&-0.14033&0.43815&$2.540 \times 10^{-2}$\cr
0.4  &0.44449&-0.30614&0.46113&0.57393&-0.04057&0.40624&$2.234 \times 10^{-2}$\cr
0.5  &0.41591&-0.22160&0.53091&0.55857&-0.22518&0.36516&$1.964 \times 10^{-2}$\cr
0.6  &0.37799&-0.26409&0.53974&0.56726&-0.26749&0.32050&$1.648 \times 10^{-2}$\cr
0.7  &0.33452&-0.30791&0.54518&0.57314&-0.31105&0.26614&$1.317 \times 10^{-2}$\cr
0.8  &0.28247&-0.35464&0.54622&0.57594&-0.35749&0.19123&$9.600 \times 10^{-3}$\cr
\hline
\end{tabular}
\caption{ A particular solution of  system (\ref{eq1},\ref{eq2},\ref{eq4},\ref{eq6}) with the accuracy $10^{-5}$ for different $p$,  
Parameters ${\cal{P}}$ from Family III 
have been set equal to zero  \label{Table01}. 
The last column shows the  
discrepancy $\delta(\rho^R)$ between the created state and the Werner state. Notice that setting the parameters from Family III
  equal to zero leads to significant changes in the results, as
  compared to Table \ref{Table0}. This happens because we deal with  particular solutions in both cases.
}
\end{table}

\section{Effect of imperfections in the chain}
\label{Section:imperfections}
Considering a set of chains of equal length governed by the same Hamiltonian $H$ we can expect minor 
differences between them owing to the unavoidable imperfections in the preparation of these chains. Consequently, the parameters 
${\cal{P}}$ (\ref{param}) vary slightly passing from one chain to another. 
Although the parameters ${\cal{P}}$  can be defined for any of these  chains using the protocol of
 Sec.\ref{Section:dp}, 
the procedure is rather complicated. Instead we can  calculate 
the parameters ${\cal{P}}$ for one of these chains and take into account the possible variations of these parameters 
in working with other chains. In this way, we, {{}generally speaking,}
 prepare an approximate state instead of the exactly required one. 
However, this can be done if the approximate state  is acceptable for our purposes.  

Thus, below  we assume that  only the two pairs of the coupling  constants 
$\delta_1$ and $\delta_2$ close to the ends of the chain in Hamiltonian (\ref{XY})  are  perfectly adjusted. 
All other coupling constants are only manufactured with a certain 
accuracy $\varepsilon$, i.e.
\begin{eqnarray}\label{perm}
D_i = 1 + \varepsilon \Delta_i ,\,\,i=3,4,\dots,N-3,
\end{eqnarray}
where $\varepsilon$ is the amplitude of random perturbations and  $\Delta_i $ 
are random numbers in the interval $-1\le \Delta_i\le 1$.
Below we present results for the randomness amplitudes $\varepsilon
=0.025$ and 
 $\varepsilon =0.05$. 

{{}
\subsection{Deviations of parameters  (\ref{param}) due to  imperfections of the Hamiltonian }}

To characterize the effect of  random imperfections of the Hamiltonian, 
we consider $N_p$ independent chains of the same  type   and characterize the  parameters 
${\cal{P}}$ associated with these chains  by their mean values and
standard deviations.
{{}Thus, for some parameter $P\in {\cal{P}}$,  the mean value  
$\big\langle P \big\rangle$ and standard deviation $\sigma(P)$ 
are calculated by the usual formulas:
\begin{eqnarray}\label{mean}
\big\langle P(\varepsilon) \big\rangle =\frac{1}{N_p}\sum_{i=1}^{N_p} P_i(\varepsilon) ,
\end{eqnarray}
\begin{eqnarray}\label{deviation}
\sigma(P(\varepsilon) ) = \sqrt{ 
\frac{1}{N_p-1}\sum_{i=1}^{N_p} \big(P_i(\varepsilon)  - \big\langle P(\varepsilon)\big\rangle\big)^2} ,
\end{eqnarray}
where $P_i$ is the value of $P$ for the $i$th chain. 
}

\begin{figure*}
\hspace*{-0.3cm}  \epsfig{file=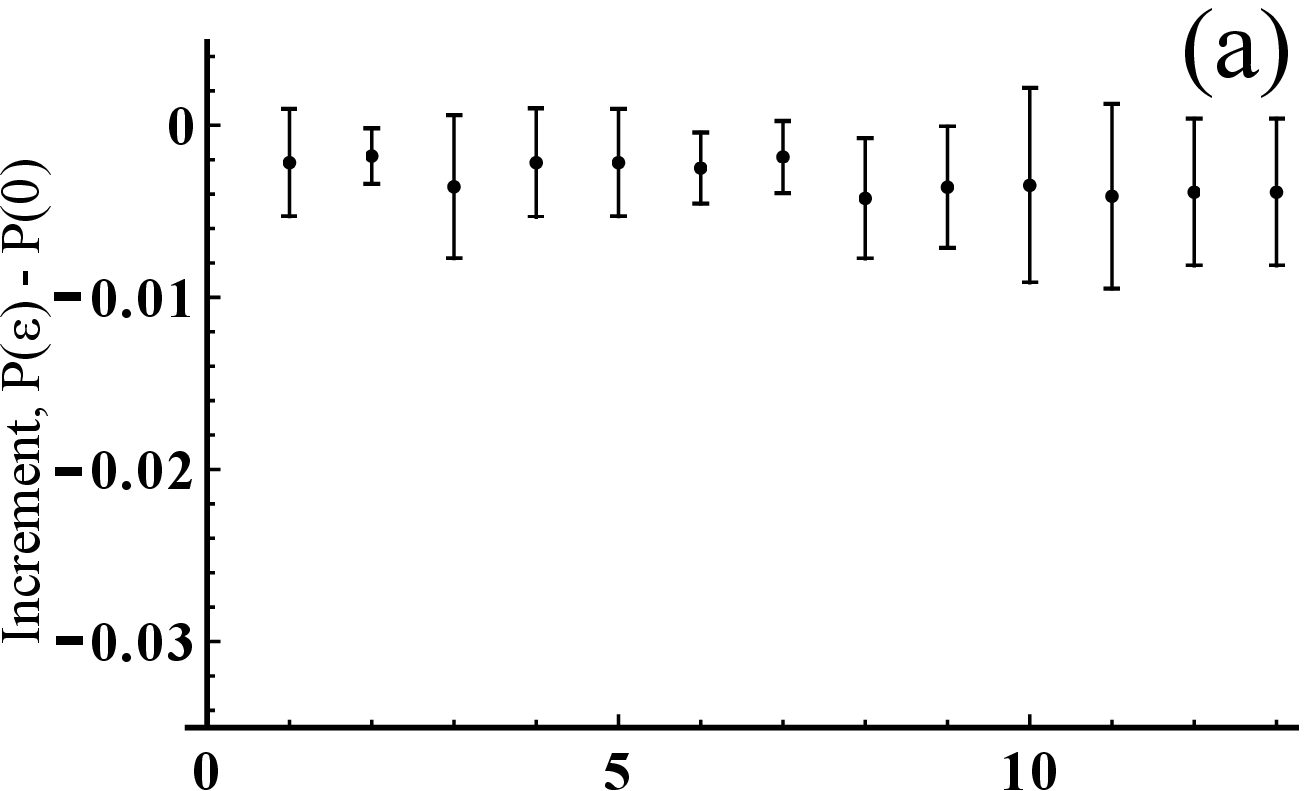, 
  scale=0.5
   ,angle=0
}  
\epsfig{file=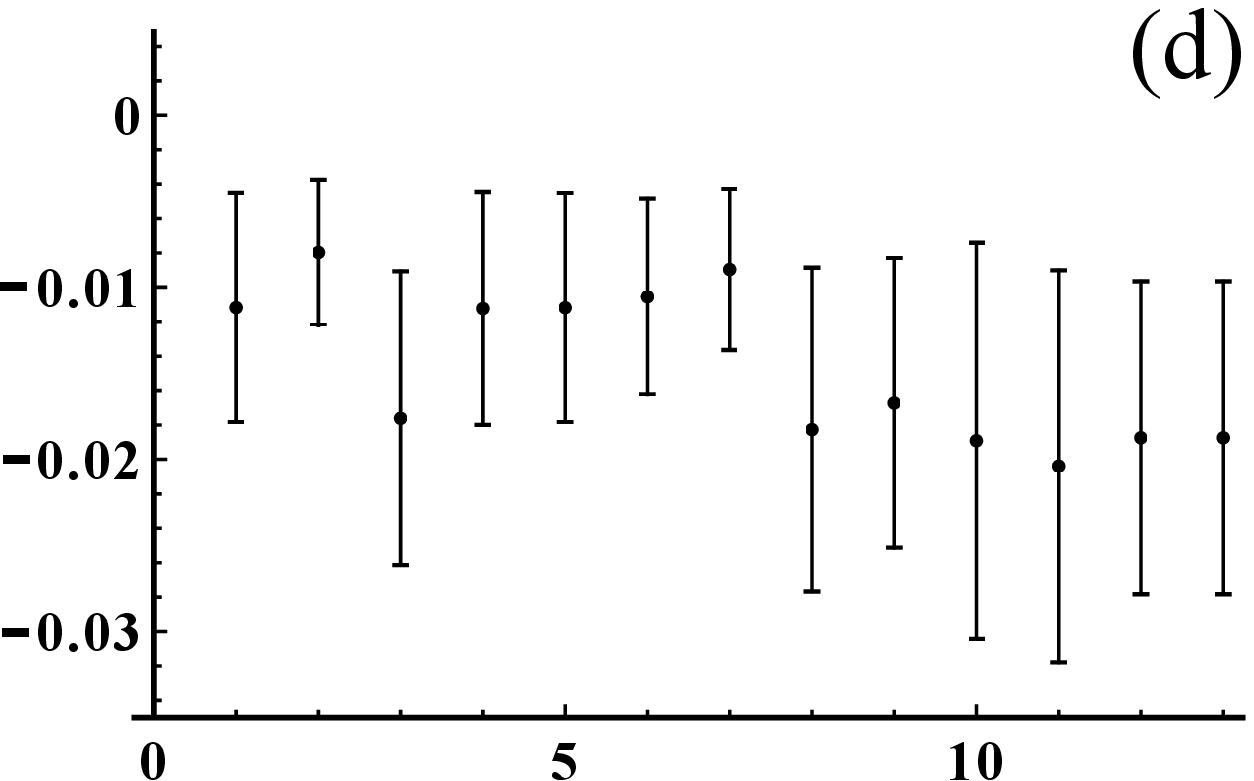, 
  scale=0.5
   ,angle=0
} \newline
\epsfig{file=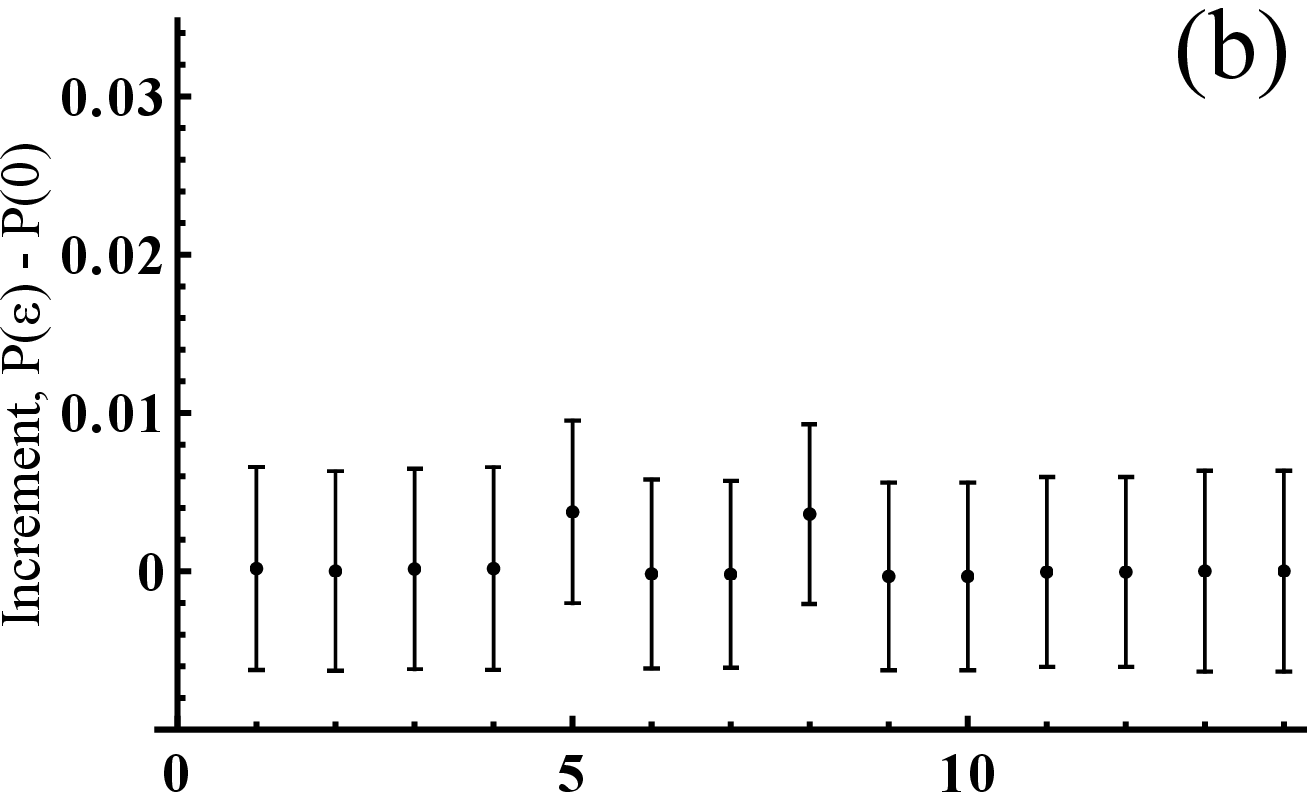, 
  scale=0.5
   ,angle=0
}
\epsfig{file=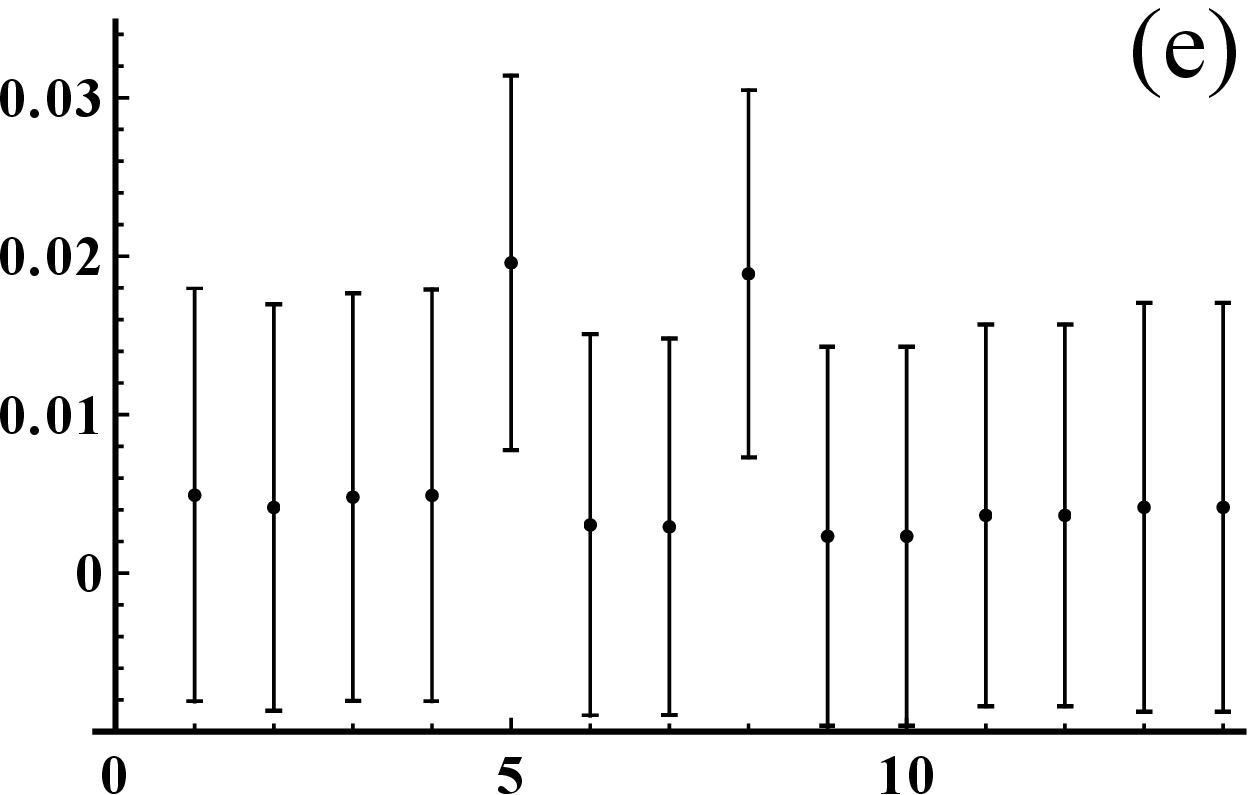, 
  scale=0.5
   ,angle=0
}\newline\hspace*{-1.2cm}
\epsfig{file=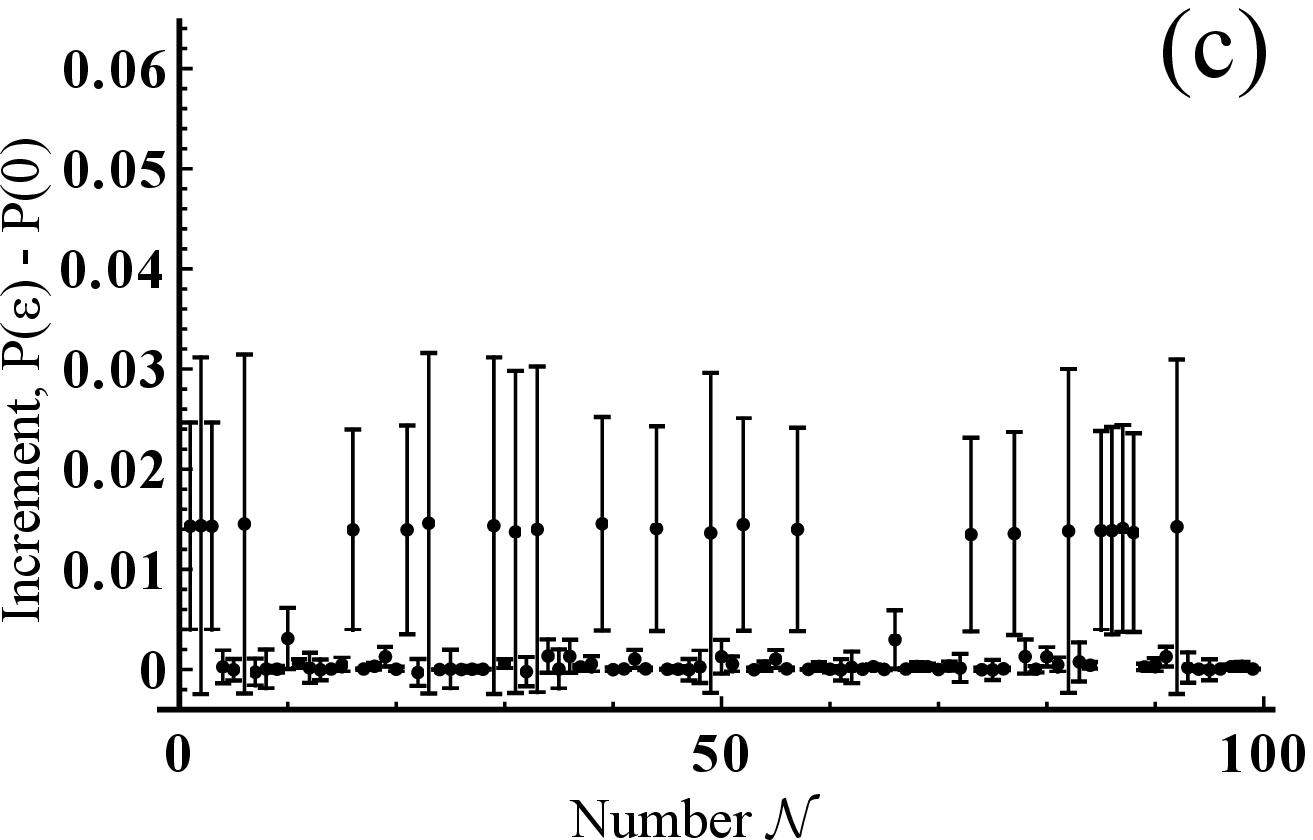, 
  scale=0.5
   ,angle=0
}
\epsfig{file=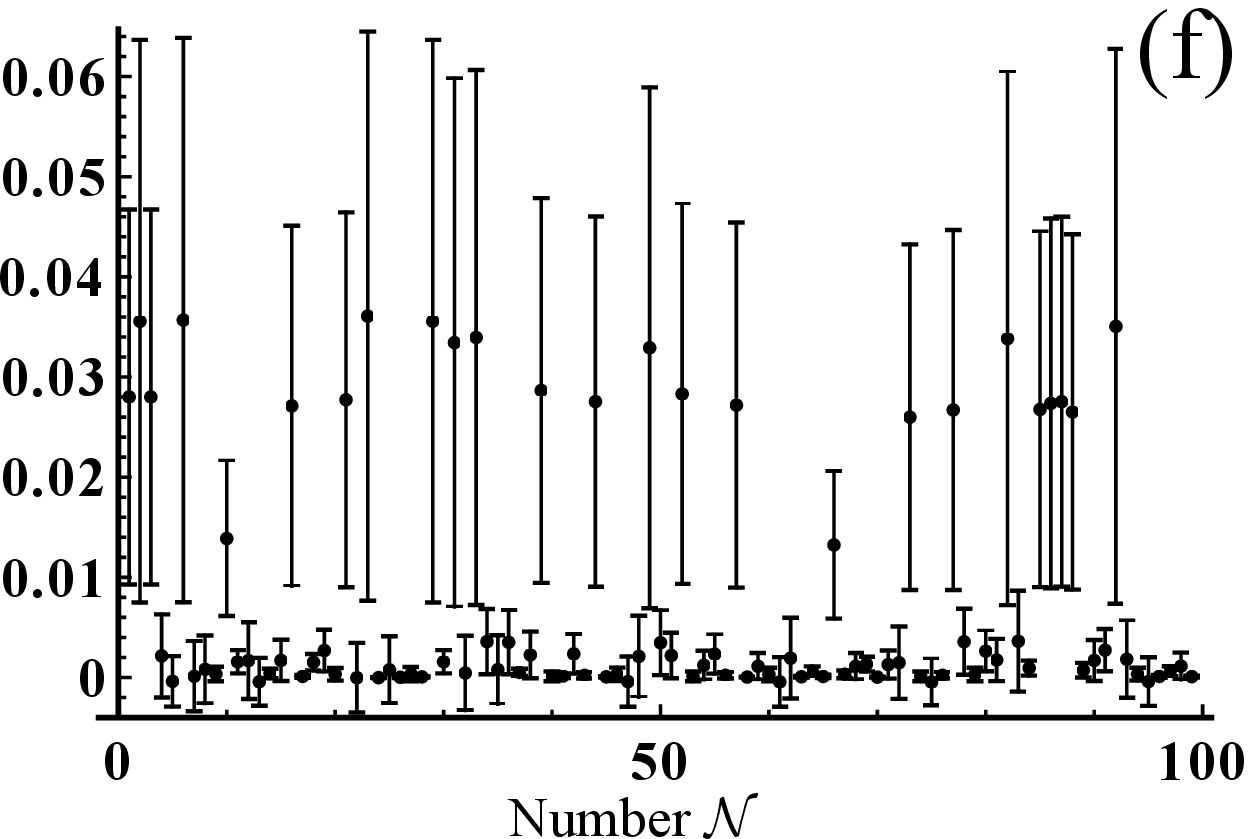, 
  scale=0.5
   ,angle=0
}
\caption{{{}Quantity $P(\varepsilon) - P(0)$ with  $P(\varepsilon)$ given in 
 (\ref{avdsp}) for three families of the parameters ${\cal{P}}$  
and two values of $\varepsilon$. {{}Averaging is over
  $N_p=100$ random chains. 
Numbers ${\cal{N}}$ enumerate} the parameters  ${\cal{P}}$
in accordance with Tables \ref{Table1}, \ref{Table21} and Appendix. Notice that the scales are different in 
these figures. {{}The mean values $\big\langle P(\varepsilon) \big\rangle$ 
are shown 
by the bold dots, the vertical bars indicate the standard deviations  
$\sigma(P(\varepsilon))$.}
(a,b,c) $ \varepsilon = 0.025$; (d,e,f) $ \varepsilon = 0.05$. 
(a,d) Family I, (b,e) Family II, (c,f) Family III {{} of parameters ${\cal{P}}$}. 
} }
  \label{Fig:avr} 
\end{figure*}

As a result, the value $P(\varepsilon)$ can be approximated as
\begin{eqnarray}\label{avdsp}
P(\varepsilon) =\big\langle P(\varepsilon) \big\rangle \pm\sigma(P(\varepsilon) ).
\end{eqnarray}
The quantities $P(\varepsilon)- P(0)$ 
{{}($P(0)$ denotes parameter obtained in 
Sec.\ref{Section:param} for the unperturbed Hamiltonian)} are represented
in Fig.\ref{Fig:avr} for the 
three families I, II, and III of
parameters ${\cal{P}}$ and for the two values of the randomness 
amplitude $\varepsilon = 0.025$, $0.05$,  averaging
  over $N_p=100$ chains.
\noindent
Fig.\ref{Fig:avr} shows  that the mean values 
{{}$\big\langle P(\varepsilon) \big\rangle$
  of the parameters from the Family I 
are less than the ideal values $P(0)$} of these parameters,  while
the mean values of the parameters 
from the Family II almost coincide
 with their ideal values. The parameters from the Family III are separated into two parts: the
 mean values of  some of them almost coincide with their
ideal values, while  the  mean values of others differ from the ideal values rather significantly.
 Fig.\ref{Fig:avr} shows also that the deviation from the calculated mean values is 
almost uniform  for the parameters from the Family II. We may conclude, 
that the imperfections of $5\%$ ($\varepsilon = 0.05$) lead to reasonable 
 approximations of the parameters  ${\cal{P}}$  with roughly the same  accuracy.

\subsection{Creation of the Werner state using a  chain of 20 nodes  with imperfections}
\label{Section:Wappr}

\begin{figure*}
   \epsfig{file=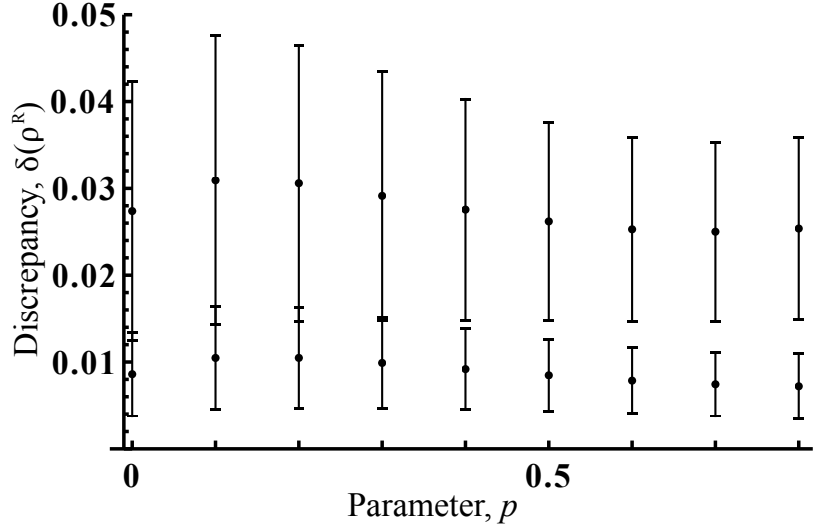,
  scale=0.8
   ,angle=0
}  
\caption{{{}{{}
Density matrix discrepancy $\delta(\rho^R(\varepsilon))= 
\big\langle \delta (\rho^R(\varepsilon))\rangle \pm 
 \sigma(\delta(\rho^R(\varepsilon))) $ for different values of the
 parameter $p$ of the Werner state.  
The averaging is over $N_p=100$ random chains.
The mean values $\big\langle \delta (\rho^R(\varepsilon))\rangle$ 
are shown 
by the bold dots, the vertical bars indicate the standard deviations  
$\sigma(\delta(\rho^R(\varepsilon)))$.
The upper and lower sets of dots correspond to
$\varepsilon =0.05$ and $\varepsilon =0.025$, respectively.}
} }
  \label{Fig:wapr} 
\end{figure*}

The parameters $a$ calculated in Sec.\ref{Section:Werner}
were determined for
 the chain with unperturbed 
 coupling constants. In this section
we use {{}these parameters} to create the Werner state using the chain of 20 nodes 
 with imperfections and characterize 
the effect of 
imperfections by the mean value of the density matrix discrepancy,
 $\big\langle \delta(\rho^R(\varepsilon)) \big\rangle$, 
and its standard deviation $\sigma(\delta(\rho^R(\varepsilon))) $. Results of our calculations are shown in 
Fig.\ref{Fig:wapr}.
{{}
From this figure we see that  the imperfection $\varepsilon =0.025$ yields discrepancy 
$\delta\lesssim 0.02$ and  imperfection $\varepsilon =0.05$ yields discrepancy 
$\delta\lesssim 0.05$
 (compare with the discrepancy  $\delta$ in the
unperturbed case,  
Table \ref{Table0}), i.e. the considered values of imperfection strength yield a reasonable 
approximation of the density matrix.  }
This figure shows also that both the mean value $\big\langle \delta \big\rangle$ and the 
standard deviation $\sigma(\delta) $ depend only slightly on the parameter $p$
 of the Werner state, {{}i.e., the  effect of imperfections
   of the Hamiltonian  on the realization of
  Werner states is almost independent of $p$. }

 \section{Conclusions}
 \label{Section:conclusions}

We consider the problem of remote creation of a two-qubit state 
using a two-qubit excitation initial state of the sender. 
 The number of qubits of the sender is not important for the protocol, although it affects  the 
creatable region of the receiver's state space. 
We show that a given communication line with sender and receiver of
fixed size  has a
certain number of parameters ${\cal{P}}$ (\ref{param}) which completely characterize the 
Hamiltonian dependence and evolution of the receiver's state. They are attributes of the
 communication line and  do not change 
during the operation. These parameters are responsible for the 
volume of the creatable region of the receiver's state-space. 
Another set of parameters characterizes the sender's initial state. This set is Hamiltonian independent and 
is referred to as a set of control parameters (\ref{contrpar}). 
These are the parameters  responsible for {{} creation of}  a particular  state from the creatable region.
 
When dealing with a particular communication line, we can define the parameters ${\cal{P}}$ using the analytical formulas 
of Sec.\ref{Section:receiver}. However, these formulas may give improper results if the 
actual  parameters of the chain differ from their theoretically prescribed values. Thus we recommend to
 define these parameters 
 performing  a set of 
remote state-creation processes with prescribed initial  conditions (the direct problem), 
Sec.\ref{Section:dp}. These parameters can be defined only with some accuracy, which is a reason of dealing with the 
approximate remote state creation in reality. Another reason is that the analytically 
calculated control parameters can not be implemented with  absolute accuracy. 
Thus, dealing with  approximately created states we have to estimate the accuracy of state creation. 
For this purpose the density matrix discrepancy $\delta(\rho^R)$
is introduced in Sec.\ref{Section:creation}. 

{{}Not all parameters ${\cal{P}}$ are equally important. In our examples of $N=20$ and $60$ nodes
 (Sec.\ref{Section:param}) we separate them into three families by their absolute values. The most significant is the first family of 13 parameters.
These are the parameters which remain (and are equal to unity) in the 
chain engineered for perfect state transfer, while all  other parameters vanish in this chain.}

Since determining the  parameters ${\cal{P}}$ for any particular chain is a rather complicated procedure, 
we may not need to calculate them separately for all chains of the same type 
(i.e. governed by the same Hamiltonian and having the same length). 
Instead, we calculate them for one of these chains, then  the effect
of  random  imperfections of the Hamiltonian
can be reflected  in the accuracy  of the state creation, if only this accuracy is acceptable. 
We estimate the  effect of imperfections on some quantity $x$ 
by  its mean value 
$\big\langle x \big\rangle$  and standard deviation $\sigma(x)$. 

All these theoretical arguments have been justified by the example of  
remote creation of the two-qubit  
Werner state with parameter $p$ in the interval $0\le p \le 0.8744$ in Secs.\ref{Section:Werner} and
\ref{Section:Wappr} using the chain of 20 nodes and two amplitudes of the Hamiltonian imperfection 
(deviation of coupling constants) 
$\varepsilon=0.025$, $0.05$.

This work is partially supported by the program of RAS 
''Element base of quantum computers'',  by the Russian Foundation for Basic Research,
 grants No.15-07-07928 and by DAAD (the Funding program "Research stay for University Academics and 
Scientists", 2015 (50015559)).
.

  \section{Appendix. Family III of  parameters (\ref{param}) in model 
of Sec. \ref{Section:model} with absolute values $< 0.02$, $N=20$}
\label{Section:Appendix}

\begin{minipage}{5cm}
\begin{tabular}{|c|c|c|}
\hline
$P\in {\cal{P}}$&${\cal{N}}$&$N= 20$\cr
\hline
$p_{19;1} $&1&$ -1.007 \times 10^{-4}$\cr
$p_{19;3} $&2&$ -7.168 \times 10^{-3}$\cr 
$p_{20;2} $&3&$ -1.007 \times 10^{-4}$\cr 
$p_{20;3} $&4&$   -1.928  \times 10^{-2}i$\cr
$p_{20;4} $&5&$ -3.860\times 10^{-3}$ \cr
$p_{19, 20;1, 3} $&6&$ 7.142\times 10^{-3} i$\cr
 $p_{19, 20;2, 3} $&7&$ 1.865\times 10^{-2}$ \cr
 $p_{19, 20;2, 4} $&8&$   - 3.743\times 10^{-3} i$\cr 
 $p_{19, 20;3, 4} $&9&$ 1.749\times 10^{-3}$  \cr
 $P_{19;1, 1, 2} $&10&$   - 7.606\times 10^{-3} i$ \cr
 $P_{19;2, 1, 2} $&11&$ 3.665 \times 10^{-6}$\cr
 $P_{19;3, 1, 2} $&12& $  - 1.858\times 10^{-2}i$  \cr
 $P_{19;4, 1, 2} $&13&$ -3.720\times 10^{-3}$ \cr
 $P_{19;1, 1, 3} $&14&$ -5.442 \times 10^{-5}$\cr
 $P_{19;2, 1, 3} $&15&$    7.194 \times 10^{- 7} i$ \cr
 $P_{19;3, 1, 3} $&16&$ -3.695\times 10^{- 5}$\cr
 $P_{19;4, 1, 3} $&17&$    2.757\times 10^{- 5}i$\cr 
 $P_{19;1, 1, 4} $&18& $ 7.023\times 10^{- 4} i$ \cr
 $P_{19;2, 1, 4} $&19& $ 8.958\times 10^{- 6}$\cr 
 $P_{19;3, 1, 4} $&20&$   1.714 \times 10^{- 3}i$ \cr
 $P_{19;4, 1, 4} $&21&$ 4.440\times 10^{- 4}$\cr
 $P_{19;1, 2, 3} $&22&$    1.858\times 10^{- 2} i$\cr
 $P_{19;2, 2, 3} $&23&$ -7.170\times 10^{- 3}$\cr
 $P_{19;4, 2, 3} $&24&$ -7.199\times 10^{-5}$ \cr
\hline
\end{tabular}
\end{minipage}
\begin{minipage}{5cm}
\begin{tabular}{|c|c|c|}
\hline
$P\in {\cal{P}}$&${\cal{N}}$&$N= 20$\cr
\hline
 $P_{19;1, 2, 4} $&25& $-3.729\times 10^{-3}$ \cr
 $P_{19;3, 2, 4} $&26&$ 7.216\times 10^{-5}$ \cr 
 $P_{19;1, 3, 4} $&27&$ 1.742\times 10^{-3}i$ \cr
 $P_{19;2, 3, 4} $&28&$ -1.762 \times 10^{- 7}$ \cr
 $P_{19;4, 3, 4} $&29&$ 7.161 \times 10^{- 3}$\cr
 $P_{20;1, 1, 2} $&30&$ -3.665\times 10^{-6}$\cr 
$P_{20;3, 1, 2} $&31&$ 6.907\times 10^{-3}$ \cr 
 $P_{20;1, 1, 3} $&32&$    1.928 \times 10^{-2}i$ \cr
 $P_{20;2, 1, 3} $&33&$ -6.909  \times 10^{-3}$\cr
 $P_{20;4, 1, 3} $&34&$ 6.377\times 10^{-4}$ \cr
$P_{20;1, 1, 4} $&35& $-3.869\times 10^{-3}$\cr
$P_{20;3, 1, 4} $&36&$ -6.375 \times 10^{-4}$\cr
 $P_{20;1, 2, 3} $&37&$ 1.878 \times 10^{- 6}$\cr 
 $P_{20;2, 2, 3} $&38&$    1.236 \times 10^{- 3} i$ \cr
 $P_{20;3, 2, 3} $&39&$ 2.344 \times 10^{- 4}$\cr
 $P_{20;4, 2, 3} $&40&$    1.665 \times 10^{-3} i$ \cr
 $P_{20;1, 2, 4} $&41&$    3.771\times 10^{- 7} i$ \cr
 $P_{20;2, 2, 4} $&42&$ -2.387 \times 10^{-4}$ \cr
 $P_{20;3, 2, 4} $&43&$    2.683  \times 10^{-5}i$\cr 
 $P_{20;4, 2, 4} $&44&$-2.335 \times 10^{-4}$\cr
 $P_{20;1, 3, 4} $&45&$1.762 \times 10^{- 7}$ \cr
 $P_{20;2, 3, 4} $&46&$   - 1.692\times 10^{- 3}i$ \cr
 $P_{20;3, 3, 4} $&47&$ -3.848\times 10^{- 3}$ \cr
$P_{20;4, 3, 4} $&48& $  - 1.912\times 10^{- 2} i$\cr
\hline
\end{tabular}
\end{minipage}
\begin{minipage}{5cm}
\begin{tabular}{|c|c|c|}\hline
$P\in {\cal{P}}$&${\cal{N}}$&$N= 20$\cr
\hline
 $P_{20, 20;1, 2, 1, 3} $&49&$    6.880\times 10^{- 3} i$\cr
 $P_{20, 20;1, 3, 1, 4} $&50&$    7.096 \times 10^{-4}i$ \cr
 $P_{20, 20;1, 2, 2, 3} $&51&$ 1.231\times 10^{- 3}$\cr
 $P_{20, 20;1, 3, 2, 3} $&52&$   - 2.335\times 10^{- 4} i$\cr 
 $P_{20, 20;1, 4, 2, 3} $&53&$ 1.659\times 10^{- 3}$ \cr
 $P_{20, 20;2, 3, 2, 3} $&54&$ 2.385\times 10^{- 5}$ \cr
  $P_{20, 20;1, 2, 2, 4} $&55&$   2.377 \times 10^{- 4}i$\cr
 $P_{20, 20;1, 3, 2, 4} $&56& $2.673\times 10^{- 5}$\cr
 $P_{20, 20;1, 4, 2, 4} $&57&$   2.326\times 10^{- 4} i$\cr
 $P_{20, 20;2, 3, 2, 4} $&58&$    4.604 \times 10^{- 6} i $\cr 
 $P_{20, 20;2, 4, 2, 4} $&59&$ 8.978 \times 10^{- 7} $ \cr
 $P_{20, 20;1, 2, 3, 4} $&60&$ -1.685\times 10^{- 3} $ \cr
 $P_{20, 20;1, 3, 3, 4} $&61&$   3.832\times 10^{- 3} i $  \cr
 $P_{20, 20;1, 4, 3, 4} $&62& $-1.905\times 10^{- 2} $ \cr
 $P_{20, 20;2, 3, 3, 4} $&63&$ -3.300\times 10^{-5} $\cr
 $P_{20, 20;2, 4, 3, 4} $&64& $  4.605  \times 10^{- 6}  i$ \cr
 $P_{20, 20;3, 4, 3, 4} $&65&$ 3.835\times 10^{-4} $\cr
 $P_{19, 19;1, 2, 1, 2} $&66&$ 7.358 \times 10^{-3} $\cr
 $P_{19, 19;1, 2, 1, 3} $&67&$   - 5.265 \times 10^{-5}i  $\cr
 $P_{19, 19;1, 3, 1, 3} $&68&$ 3.864 \times 10^{- 7}$ \cr
 $P_{19, 19;1, 2, 1, 4} $&69&$ -6.795\times 10^{-4}$  \cr
 $P_{19, 19;1, 3, 1, 4} $&70&$   - 4.862 \times 10^{- 6}i $  \cr
$P_{19, 19;1, 4, 1, 4} $&71&$ 6.275\times 10^{- 5} $  \cr
 $P_{19, 19;1, 2, 2, 3} $&72&$ -1.797\times 10^{- 2} $\cr
 \hline
\end{tabular}
\end{minipage}
\begin{minipage}{8cm}
\begin{tabular}{|c|c|c|}\hline
$P\in {\cal{P}}$&${\cal{N}}$&$N= 20$\cr
\hline
 $P_{19, 19;1, 3, 2, 3} $&73&$   -3.574 \times 10^{-5} i$\cr 
 $P_{19, 19;1, 4, 2, 3} $&74&$ 1.659\times 10^{-3} $ \cr
 $P_{19, 19;1, 2, 2, 4} $&75& $  - 3.598 \times 10^{-3}  i$ \cr
 $P_{19, 19;1, 3, 2, 4} $&76& $2.673 \times 10^{-5} $\cr
 $P_{19, 19;1, 4, 2, 4} $&77& $  4.304  \times 10^{-4}i $\cr
 $P_{19, 19;2, 3, 2, 4} $&78& $  - 7.098  \times 10^{-4}i$\cr 
 $P_{19, 19;1, 2, 3, 4} $&79& $-1.685 \times 10^{-3}$\cr
 $P_{19, 19;1, 3, 3, 4} $&80& $  -3.497\times 10^{-6} i$\cr 
 $P_{19, 19;1, 4, 3, 4} $&81& $1.563 \times 10^{-4}$ \cr
 $P_{19, 19;2, 4, 3, 4} $&82& $  - 6.928\times 10^{-3}  i$ \cr
 $P_{19, 19;3, 4, 3, 4} $&83& $8.021\times 10^{-3}$\cr
 $P_{19, 20;1, 2, 1, 2} $&84& $  2.884 \times 10^{-6}i$\cr 
$P_{19, 20;1, 3, 1, 3} $&85&  $-3.785\times 10^{-5}  i$ \cr
$P_{19, 20;1, 4, 1, 4} $&86&  $  4.450\times 10^{-4}i$ \cr
 \hline
\end{tabular}
\end{minipage}
\begin{minipage}{8cm}
\begin{tabular}{|c|c|c|}\hline
$P\in {\cal{P}}$&${\cal{N}}$&$N= 20$\cr
\hline
$P_{19, 20;2, 3, 2, 3} $&87& $  - 2.356\times 10^{-4} i $\cr
 $P_{19, 20;2, 4, 2, 4} $&88&$   2.472\times 10^{- 4}  i$ \cr
$P_{19, 20;3, 4, 3, 4} $&89&$  2.065\times 10^{-4} i$  \cr
$P_{19, 20;1, 3, 1, 2}=P_{19, 20;3,4,2,4} $&90&$ 7.220 \times 10^{- 7} $ \cr
 $P_{19, 20;1, 4, 1, 2}=P_{19, 20;3,4,2,3}^*$&91&$   8.994\times 10^{- 6} i $ \cr
$P_{19, 20;2, 3, 1, 2}=P_{19, 20;3,4,1,4}^* $&92&$   - 7.140\times 10^{-3} i$ \cr
$P_{19, 20;1, 2, 1, 3}=P_{19, 20;2,4,3,4} $&93&$  -1.865\times 10^{-2} $ \cr
$P_{19, 20;1, 4, 1, 3}=P_{19, 20;2,4,2,3} $&94&$ 1.721\times 10^{-3} $\cr
 $P_{19, 20;1, 2, 1, 4}=P_{19, 20;2,3,3,4}^* $&95& $  - 3.734 \times 10^{-3}i$ \cr
 $P_{19, 20;1, 3, 1, 4}=P_{19, 20;2,3,2,4} $&96&$ 2.767\times 10^{-5}$ \cr
 $P_{19, 20;1, 2, 2, 3}=P_{19, 20;1,4,3,4}^* $&97& $1.942 \times 10^{- 6}  i$ \cr
$P_{19, 20;1, 3, 2, 3}=P_{19, 20;1,4,2,4} $&98& $1.015 \times 10^{- 8} $ \cr
 $P_{19, 20;1, 2, 2, 4}=P_{19, 20;1,3,3,4} $&99 &$ -3.888 \times 10^{- 7} $ \cr
 \hline
\end{tabular}
\end{minipage}

\vspace{0.5cm}
{{}We also observe a symmetry among the parameters $P_{12,20;klnm}$
in lines ${\cal{N}}=90,\dots,99$ of the above  Table. This is an ``approximate`` symmetry which is destroyed 
in higher order approximation. In addition, $P_{19, 20;klnm} =0 $,
$k\neq l\neq n\neq m$. Both of these facts  
appear due to the assumption of nearest-neighbor couplings in Hamiltonian (\ref{XY})  and the mirror symmetry of the chain. }

\end{document}